\documentclass[8.5pt,twoside,twocolumn]{article}
\usepackage[super,sort&compress,comma]{natbib} 
\usepackage{mhchem}
\usepackage{times,mathptmx}
\usepackage{sectsty}
\usepackage{balance} 

\usepackage{graphicx} 
\usepackage{lastpage}
\usepackage[format=plain,justification=raggedright,singlelinecheck=false,font=small,labelfont=bf,labelsep=space]{caption} 

\newcommand\blfootnote[1]{%
  \begingroup
  \renewcommand\thefootnote{}\footnote{#1}%
  \addtocounter{footnote}{-1}%
  \endgroup
}

\begin{document}

\thispagestyle{plain}

\twocolumn[
  \begin{@twocolumnfalse}
\noindent\LARGE{\textbf{Folding kinetics of a polymer [corrigendum]}}
\vspace{0.6cm}

\noindent\large{\textbf{Bart Vorselaars,\textit{$^{a}$}\textit{$^{\!,b}$} \v{S}t\v{e}p\'{a}n R\r{u}\v{z}i\v{c}ka,\textit{$^{a}$} David Quigley,\textit{$^{a}$} and
Michael P. Allen\textit{$^{a}$}\textit{$^{\!,c}$}}}\vspace{0.5cm}

\noindent \normalsize{In our original article (\textit{Phys. Chem. Chem. Phys.}, 2012, \textbf{14}, 6044–6053) a convergence problem resulted in an averaging error in computing the entropy from a set of Wang-Landau Monte-Carlo simulations. Here we report corrected results for the freezing temperature of the homopolymer chain as a function of the range of the non-bonded interaction. We find that the previously reported forward-flux sampling (FFS) and brute-force (BF) simulation results are in agreement with the revised Wang-Landau (WL) calculations. This confirms the utility of FFS for computing crystallisation rates in systems of this kind.}
\vspace{0.5cm}
 \end{@twocolumnfalse}
  ]

\blfootnote{\textit{$^{a}$~Department of Physics, University of Warwick, Coventry CV4 7AL, United Kingdom}}
\blfootnote{\textit{$^{b}$~E-mail: B.Vorselaars@warwick.ac.uk}}
\blfootnote{\textit{$^{c}$~E-mail: M.P.Allen@warwick.ac.uk}}

\begin{figure}
\centering
  \includegraphics[bb = 22 180 557 600,clip,width=6.6cm]{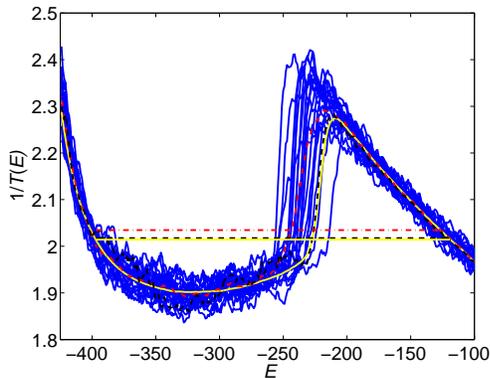}
  \caption{(Fig.~2 revised) The inverse temperature $T(E)^{-1}=\partial S(E)/\partial E$ for 20 independent WL simulations with $m=20$ iterations (blue), obtained from entropy averaging (dash-dotted red), from density-of-states averaging (dashed black) and for a simulation with $m=28$ iterations (solid yellow), all for $\chi=1.07$. Accompanying horizontal equal-area lines from a Maxwell construction have the same colour.}
  \label{fgr:invtempvariousav}
\end{figure}

We previously \citep{Ruzicka2012} found a small but significant discrepancy in the folding-unfolding transition temperature $T_f$ of a polymer chain: WL\cite{Wang2001} results were shifted with respect to both FFS \cite{Allen2006a} and BF results. Here $T_f$ for the WL results follow from an equal-area Maxwell construction of the derivative of the entropy with respect to energy, $\partial S(E)/\partial E$. The entropy was determined by averaging the results of 20 independent WL simulations $i$, $S_{\overline{S}}(E)=\langle S_i(E)\rangle$.  The derivative of the entropy for each simulation and for the average are displayed in Fig.~\ref{fgr:invtempvariousav} as solid blue lines and a dash-dotted red line, respectively. The entropy varies substantially between the runs, and hence the averaging method influences the final result. For example, one could average the density of states (DOS) $W(E)=\exp(S(E))$ instead. Here the DOS is normalized, so that its integral over all energies equals 1. The entropy of the DOS averaged over the simulations equals $S_{\overline{W}}(E)=\log \langle\exp(S_i(E))\rangle$ (its derivative is the black dashed line in Fig.~\ref{fgr:invtempvariousav}). The derivative of $S_{\overline{W}}(E)$ differs substantially from the one of $S_{\overline{S}}(E)$ with respect to the position of the upwards trend in $1/T(E)$. This also has consequences on $T_f$: the difference between the two different averages is about 1\% for $\chi=1.07$ using the Maxwell construction (horizontal lines in Fig.~\ref{fgr:invtempvariousav}).

\begin{figure}
\centering
  \includegraphics[bb = 22 180 557 600,clip,width=6.8cm]{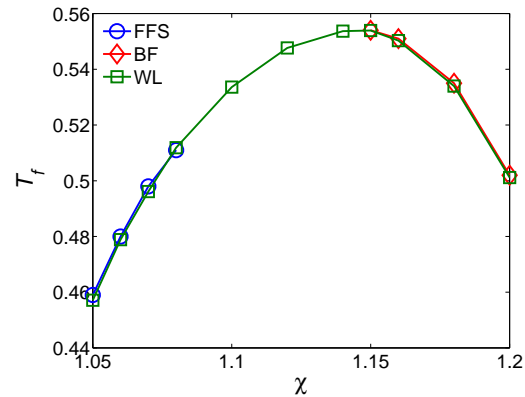}
  \caption{(Fig.~4 revised) Phase diagram of the $N=128$ bead polymer chain. Here $\chi$ is the ratio of the maximum attraction cut-off to the hard-sphere diameter, and $T_f$ the freezing temperature determined from three different methods.}
  \label{fgr:phasediagram}
\end{figure}

The noise in the WL results and the resulting influence of the average procedure has been reduced as follows. We increased the number of WL iterations from 20 to at least 28, following \citet{Taylor2009b}, and we tightened up the flatness-criterion: the next iteration is started when each entry in the `visit'-histogram is within 10\% of the average as opposed to allowing that the minimum deviates at most 20\% from the average. There are also a few other minor changes to the algorithm, to be published elsewhere. The result for $T(E)^{-1}$ from one of these improved simulations is shown in Fig.~\ref{fgr:invtempvariousav}, yellow line. Two other independent runs gave almost indistinguishable results. Observe that in the transition region, near $E=-225$, the derivative of $S_{\overline{W}}(E)$ converges faster to the extended simulations in comparison to $S_{\overline{S}}(E)$, but that in the flat region the fluctuations in the latter are smaller. Nevertheless, the transition temperature from the DOS-average is much closer to the improved results than the entropy-average: the difference is respectively 0.1\% and 1\% for $\chi=1.07$. Results for other values of $\chi$ show the same tendency. The reason that averaging the DOS gives a better result might be that in our WL simulations the entropy of the lower energies is usually underestimated before convergence. Weighting with the DOS effectively punishes these low-entropy results.
\begin{figure}
\centering
  \includegraphics[bb = 22 180 557 600,clip,width=7cm]{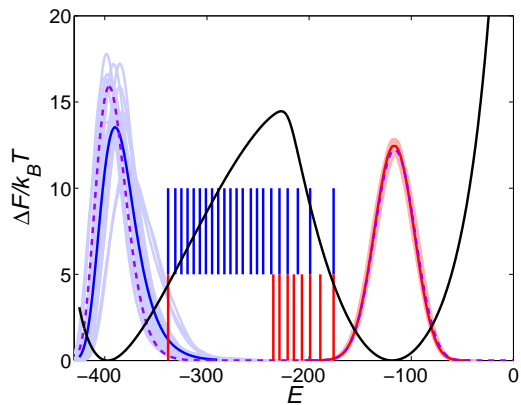}
  \caption{(Fig.~3 revised) Free energy relative to its minimum (black) at $T=T_f^{WL}=0.4964$, and energy distribution functions at $T=0.498$ for WL (purple), for the various collision dynamics runs in the folded (blue) and unfolded state (red), all for $\chi=1.07$. The respective averaged distributions are coloured darker.  The probability axis is not shown, but each distribution function has equal area. The vertical lines show the subsequent FFS interfaces for $A\to B$ (blue) and $B\to A$ (red).}
  \label{fgr:free-energy}
\end{figure}

\begin{figure}
\centering
  \includegraphics[bb =22 180 557 600, clip,width=7cm]{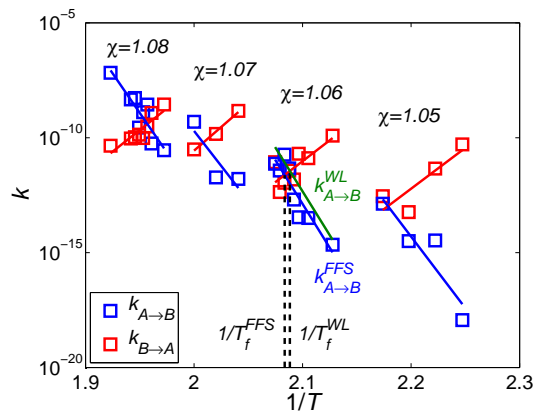}
  \caption{(Fig.~5 revised) Chevron plots for various $\chi$. The schematic for $\chi=1.06$ indicates the unfolding rate for it to bring $T_f^\mathrm{FFS}$ into agreement with $T_f^\mathrm{WL}$.}
\label{fgr:chevron}
\end{figure}

The phase diagram with the revised WL transition temperatures is given in Fig.~\ref{fgr:phasediagram}. The maximum discrepancy in $T_f$ between the FFS/BF and WL data has reduced from approximately 4\% to 0.4\%, and is now within the statistical error of the simulations. The original differences may seem small, but note that a difference in the transition temperature of approximately 2\% results in an apparent shift of one of the rate constants by more than 4 orders of magnitude, as was previously the case for $\chi=1.06$. This is caused by the relatively large energy difference between the folded $A$ and unfolded $B$ state, $E_{B}-E_{A}\approx270$ for $\chi=1.07$ (Fig.~\ref{fgr:free-energy}). As $T_f=0.4964$, a change in the transition temperature by 0.3\% corresponds to a shift in the relative occupancy of the state $A$ to $B$ by a factor $(P_{T_1}(E_{B})/P_{T_1}(E_{A}))/(P_{T_2}(E_{B})/P_{T_2}(E_{A}))\approx 5$ (where $P_T(E)\propto W(E)\exp(-E/T)$) and hence a similar change in one of the rate constants, if the reverse rate is assumed constant. This sensitivity might contribute to the observed discrepancy in the rates between simulations and experiments\citep{Auer2001}.

The updated chevron plot is given in Fig.~\ref{fgr:chevron}. For $\chi=1.06$ the apparent shift in one of the rate constants, assuming the reverse rate is exact, is indicated. It is now approximately half an order of magnitude and within the noise of the WL and FFS results (e.g.\ discarding the first red data point halves the difference). Upon decreasing the statistical noise other small inaccuracies might be revealed. For example, there is a small discrepancy between WL and FFS in the energy distribution of the folded states (purple vs dark blue line in Fig.~\ref{fgr:free-energy}). As opposed to a pure 1D potential with two minima, there are multiple folded states which can not be reached from each other within the time scale of our simulations. Therefore one has to switch to ensemble-averaging. The ensemble of configurations that we used as starting points for the crystallized $A$ states is based on the configuration with the lowest energy that we obtained at the last FFS interface in the folding direction $E_m^B$. This still gives a slightly too high average energy. An improved scheme might require the first and last FFS interface to be located closer to their respective free energy minimum, and hence knowledge of the latter should be at hand. In this respect the WL and FFS methods complement each other. Nevertheless, as can be judged from the phase diagram, this is only a minor effect.  

We conclude that within statistical error the three methods give consistent results and confirm that forward-flux sampling (FFS) is also a viable and accurate method for use in crystallisation studies of this kind.

\footnotesize{
\providecommand*{\mcitethebibliography}{\thebibliography}
\csname @ifundefined\endcsname{endmcitethebibliography}
{\let\endmcitethebibliography\endthebibliography}{}

}

\end{document}